\documentclass[prl,twocolumn,showpacs,superscriptaddress,amsmath,amssymb]{revtex4}
\usepackage{dcolumn,epsfig,times,color,graphicx,amssymb,amsmath}%
 
 \definecolor{Blue}{rgb}{0,0,1}
\definecolor{NavyBlue}{rgb}{0.14,0.14,0.56}
\definecolor{rot}{cmyk}{0,1,1,0}
\newcommand{\e}[1]{\cdot 10^{#1}}  \newcommand{\wn}{\,cm$^{-1}$}
 \newcommand{\dg}{$^{\circ}$} 
\newcommand{\bs}{\boldsymbol}

\begin{document}

\title{The two-dimensional Brillouin zone of uniaxially strained graphene}
\author{Marcel Mohr}
 \email{marcel@physik.tu-berlin.de}
\affiliation{Institut f\"ur Festk\"orperphysik, Technische Universit\"at Berlin,
Hardenbergstr. 36, 10623 Berlin, Germany}

\author{Konstantinos Papagelis }
\affiliation{ Materials Science Department, 26504 Rio, Patras}

\author{Janina Maultzsch}
\affiliation{Institut f\"ur Festk\"orperphysik, Technische Universit\"at Berlin,
Hardenbergstr. 36, 10623 Berlin, Germany}

\author{Christian Thomsen}%
\affiliation{Institut f\"ur Festk\"orperphysik, Technische Universit\"at Berlin,
Hardenbergstr. 36, 10623 Berlin, Germany}

\date{\today}

\begin{abstract}
  We present an in-depth analysis of the electronic and vibrational band structure of
  uniaxially strained graphene by \emph{ab-initio} calculations. Depending on the
  direction and amount of strain, the Fermi crossing moves away from the
  $K$-point. However, graphene remains semimetallic under small strains.  The
  deformation of the Dirac cone near the $K$-point gives rise to a broadening of the
  $2D$ Raman mode. In spite of specific changes in the electronic and vibrational band
  structure the strain-induced frequency shifts of the Raman active $E_{2g}$ and $2D$
  modes are independent of the direction of strain. Thus, the amount of strain can be
  directly determined from a single Raman measurement.
\end{abstract}


\maketitle

The discovery of graphene in 2004 has led to strong research activities in the
last years\cite{novoselov04}. Graphene has been shown to possess  unique material properties.
In graphene the quantum Hall effect could be observed at room temperature
\cite{novoselov05,zhang05}. Due to the specific band structure at the Fermi level, the
electrons can be described as massless Dirac fermions.
Thus they mimic relativistic particles with zero rest mass and
with  an effective 'speed of light' $c' \approx 10\e{6}$\,m/s
\cite{novoselov07}.
Due to expected ballistic transport  graphene is considered to serve as
building block for micro-electronics. For this the graphene sheets have to
be grown on an insulating material such as SiO$_2$ with a different lattice
constant. This introduces strain, and the effect on the electronic properties is
therefore extremely important.  Recently, 
 uniaxially strained graphene has been investigated by Raman spectroscopy\cite{ni08,mohiuddin09,huang09}. 
The amount of  strain influences the frequency of the lattice vibrations. In addition, the polarization of the Raman signal gives
information on the orientation of the graphene sample\cite{mohiuddin09,huang09}. 

Theoretically the effect of uniaxial strain on the electronic properties has been
investigated, and the opening of a band gap has been suggested \cite{ni08, gui08}. The
question as to whether a gap opens for small strains has remained under considerable
controversy\cite{ni08, ni09, pereira09}. The effect of hydrostatic and shear strain on
the vibrational properties has been discussed in Ref.\cite{thomsen02}.  Effects of uniaxial
strain on the vibrational properties have been investigated \emph{via ab
  initio}-calculations only for very large strains on the order of 40\,\% and only for
armchair and zigzag directions.\cite{liu07} In contrast, the maximum strain realized
experimentally has been a few percent\cite{ni08, ni09, pereira09}.

Here we investigate the effect of small strains along arbitrary directions on the
electronic and vibrational properties of graphene.  We demonstrate that graphene remains
semimetallic under strain and show how the Fermi crossing moves away from the
high-symmetry points.  In addition the Dirac cones become compressed.  This influences the double-resonant $2D$ Raman mode
\cite{thomsen00,maultzsch04dr} as the double-resonance (DR) condition is altered. The
induced anisotropy leads to different DR conditions for different $k$-vectors and to a
broadening of the $2D$-peak
. This broadening is on the order of 10\wn \ for 1\,\%
strain when using laser lines in the visible spectrum.
Although specific changes in the electronic and phononic band structure for different
strain directions are found, the frequencies of the Raman active $E_{2g}$ and $2D$ modes
are independent of the direction of strain. Thus the amount of strain can be directly
determined from a single Raman measurement.

\begin{figure}[htb]
  \epsfig{file=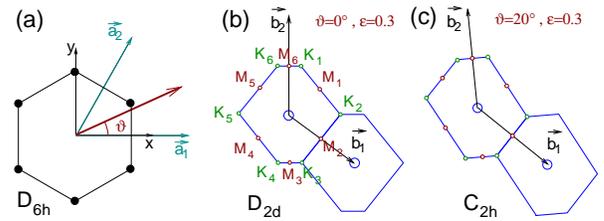,width=0.9\columnwidth}
 \caption{\label{bild:sechsecke} (Color online)  (a)
The unit cell vectors $\vec{a_1}, \vec{a_2}$ of hexagonal graphene. The arrow indicates the direction of the
applied strain. $\vartheta=0$\dg always corresponds to the zigzag direction.
(b)Brillouin zone of uniaxially strained graphene in the direction
$\vartheta=0$\dg. The point group reduces to $D_{2d}$. $\vec{b_1}, \vec{b_2}$ are the
reciprocal lattice vectors.
(c)Brillouin zone of uniaxially strained graphene in the direction
$\vartheta=20$\dg. The point group reduces to $C_{2h}$
 }
 \end{figure}

Calculations were performed with the code {\sc Quantum-Espresso}\cite{qe}.  We used a
plane-wave basis set, RRKJ pseudopotentials\cite{rappe90} and the generalized gradient
approximation in the Perdew, Burke and Ernzerhof parameterization for the
exchange-correlation functional\cite{perdew96}.  A Gaussian smearing with a width of
0.02 Ry was used.  We carefully checked the convergence in the energy differences
between different configurations and the phonon frequencies with respect to the wave
function cutoff, the charge density cutoff, the $k$-point sampling of the Brillouin
zone, the number of $q$-points for calculating the dynamical matrix and the interlayer
vacuum spacing for graphene. Energy differences are converged within 5\,meV/atom or
better, and phonon frequencies from the whole Brillouin zone within 5\,cm$^{-1}$.  The valence electrons were
expanded in a plane wave basis with an energy cutoff of 60 Ry.  A $42\times42\times1$
sampling grid was used for the integration over the Brillouin zone. The dynamical
matrices were calculated on a $12\times12\times1$ $q$-grid using the implemented
linear-response theory. Force constants were obtained via a Fourier transformation and
interpolated to obtain phonons at arbitrary points in the Brillouin zone.  All
frequencies were multiplied by a constant factor to match the experimental Raman
frequency of graphene.

\begin{figure}[htbp]
\epsfig{file=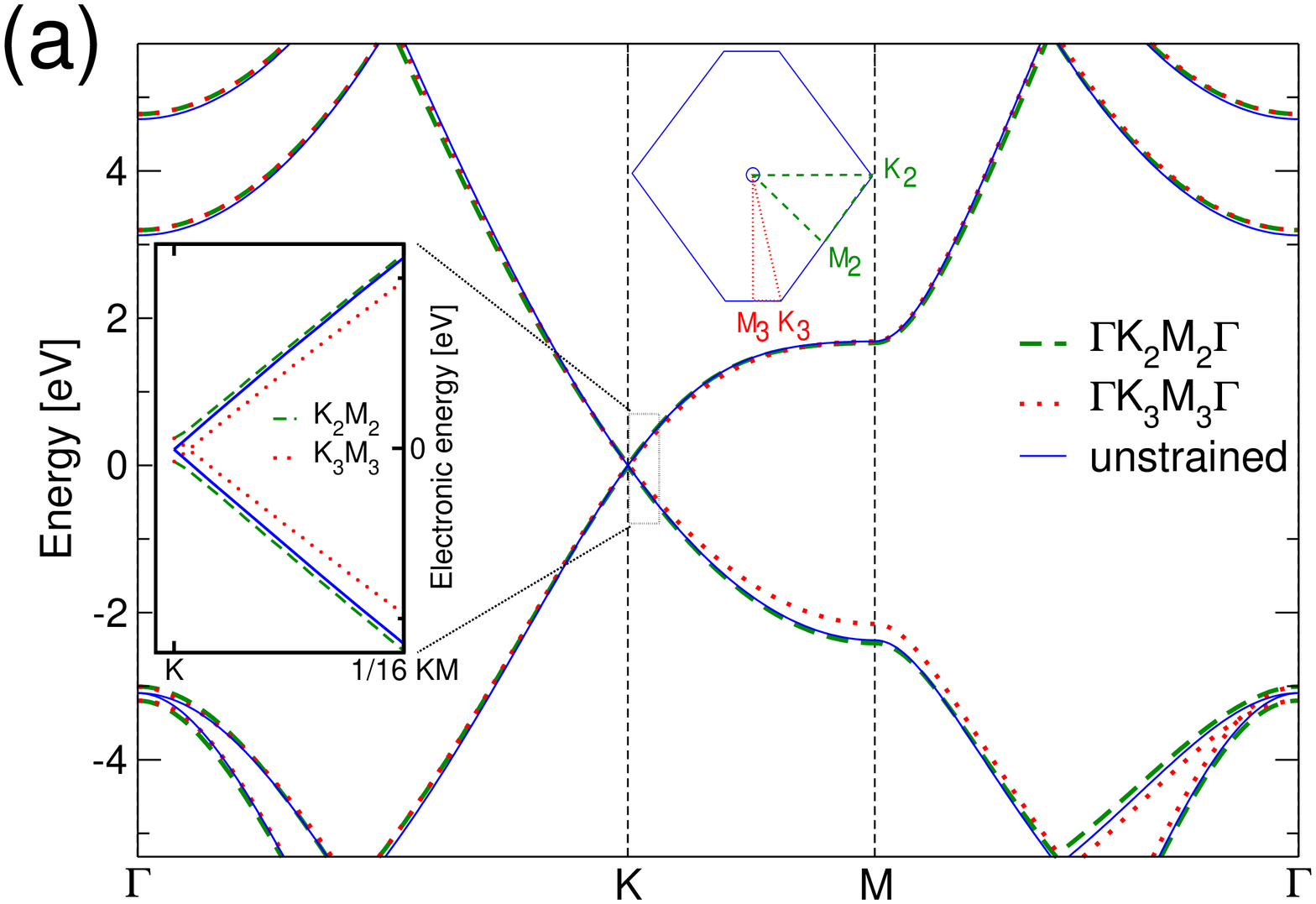,width=1.0\columnwidth}
\epsfig{file=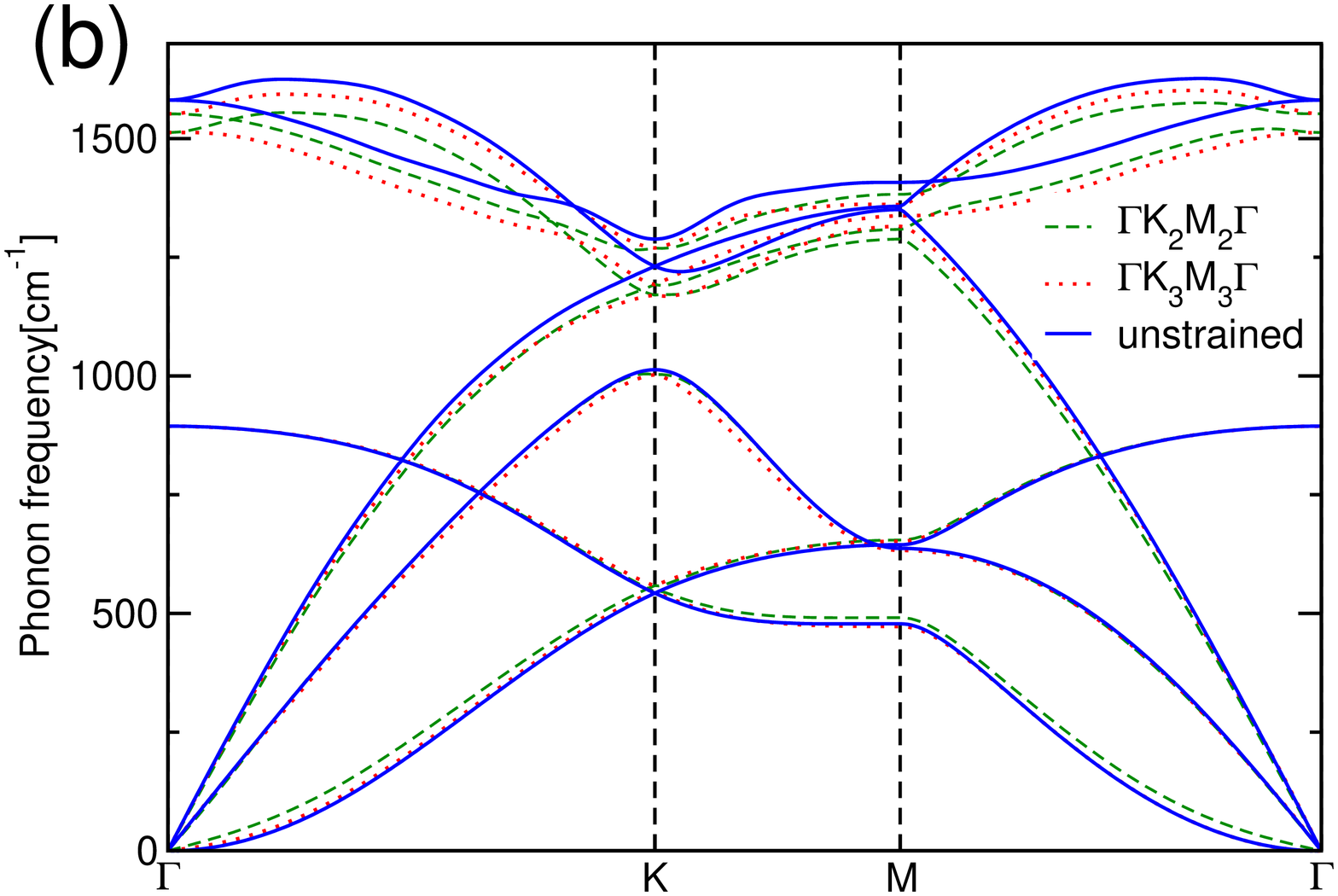,width=1.0 \columnwidth}
 \caption{\label{bild:GKMG} (Color online) (a) The electronic band structure and
(b) the phonon dispersion curves of relaxed
   and $\epsilon=0.02$-strained graphene along the $\vartheta=0$\dg-direction. The
   corresponding paths in the Brillouin zone are indicated. A closeup for the
   electronic bands along KM is  shown in the inset of (a).
 }
 \end{figure}

 The general 3-dimensional Hooke's law connects the stress tensor $\bs\sigma$ and the
 strain tensor $\bs\epsilon$ via the stiffness tensor $\bs c$:
 $\sigma_{ij}=c_{ijkl}\cdot \epsilon_{kl}$. As we are only interested in planar
 strain, the tensile strain along the $x$-axis can be expressed \emph{via} the
 two-dimensional strain tensor ${\bs\epsilon}=((\epsilon ,0),(0, -\epsilon \nu))$,
 where $\nu$ is the Poisson's ratio. Since we want to apply arbitrary strain
 directions the strain tensor has to be rotated $\bs\epsilon'=\bs R^{-1}\bs\epsilon\bs
 R$, where $\bs R$ is the rotational matrix. Here the $x$-axis corresponds to the
 zigzag direction of graphene.  After applying a finite amount of strain we relax the
 coordinates of the basis atoms until forces are below 0.001\,Ry/a.u. and minimize the
 total energy with respect to the Poisson's ratio $\nu$. For small strain values we
 obtain a Poisson's ratio $\nu$=0.164, in excellent agreement with experimental
 tension measurements on pyrolytic graphite that yield a value of
 $\nu$=0.163.\cite{blakslee70} Strain reduces the symmetry of the hexagonal
 system. For strain in arbitrary directions the point group is reduced from $D_{6h}$
 to $C_{2h}$, only the $C_2$ rotation (rotation by 180\dg ) and the inversion are
 retained (and the trivial mirror plane).  For strain along the 0\dg or 30\dg
 directions additionally mirror planes remain, resulting in $D_{2d}$-symmetry.  A
 sketch of the hexagonal lattice and the corresponding strain in real space can be
 seen in Fig.~\ref{bild:sechsecke}(a). Due to the hexagonal symmetry only strain in
 the range between $\vartheta$=0\dg and 30\dg are physically interesting.

In Figs.~\ref{bild:sechsecke} (b) and (c) the resulting Brillouin zones for strain in
the directions 0\dg and 20\dg \ are shown. For clarity we have used an exaggerated
strain of $\epsilon=0.3$.  Like in unstrained graphene, the six corner points of the
Brillouin zone correspond to the $K$-points. In the unstrained lattice these
$K$-points, e.g., $K_1$ can be expressed \emph{via} $K_1=1/3 \bs{b_1}+2/3 \bs{b_2}$, where $b_i$
are the reciprocal lattice vectors. The same definition does not hold anymore for the
strained lattice, where the $K$-points can be obtained by constructing the
perpendicular bisectors to all neighboring lattice points and determine their
intersection point.

\begin{figure*}[htb]
  \epsfig{file=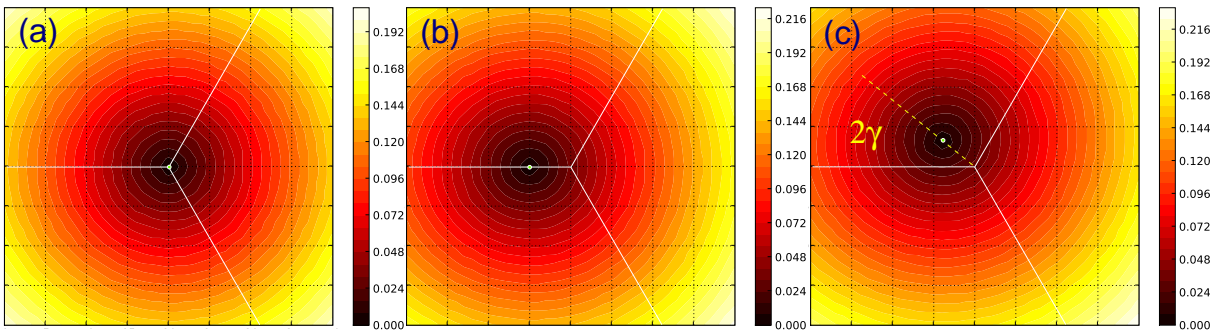,width=1.9\columnwidth}
  \caption{\label{bild:ElContour} (Color online) Contour plots of the direct optical
    transition energy of the $\pi$-$\pi^*$ -bands. The energy is given in eV. Each is
    plot centered at its respective $K$-point. The length of each side is 0.02$\pi
    /a$.  Shown here is the plot around $K_1$. Plots around $K_i$ (i=3,5) are
    identical. $K_i$ (i=2,4,6) can be found by inversion.  (a) Relaxed configuration,
    (b) 2\%-strained configuration for $\vartheta=0$\,\dg \ and (c) a 2\%-strained
    configuration for $\vartheta=20$\,\dg.  }
 \end{figure*}

 The usual way to plot band structures is along the high symmetry directions. For
 unstrained graphene this corresponds to $\Gamma -K- M- \Gamma$. In strained graphene,
 as the 6-fold symmetry is broken these lines are not sufficient to cover all
 high-symmetry lines. For strain in the $\vartheta=0$\,\dg direction we choose the
 lines $\Gamma- K_2- M_2- \Gamma$ and $\Gamma- K_3- M_3- \Gamma$, the remaining ones
 can be found by symmetry.  In Fig.~\ref{bild:GKMG} we plot the electronic band
 structure and the phonon dispersion curves for 2\% \ strain and the relaxed
 graphene. The most obvious changes in the electronic band structure include an
 increase in energy of the $\pi$-type valence bands along the $K-M$ direction. The
 splitting of the $\sigma$-type valence bands at $E=-3\,$eV can clearly be seen.  A
 closeup of the vicinity of the $K$-point reveals that the crossing of the Fermi level
 only takes place between $K_3$ and $M_3$. To further investigate the position of the
 Fermi crossing we evaluate the direct optical transition (DOT) energy of the
 $\pi$-$\pi^*$ -bands in the vicinity of the $K_1$-point.  In
 Fig.~\ref{bild:ElContour} we show energy contour plots of the DOT energy of the
 $\pi$-$\pi^*$ -bands. Each plot is centered at its respective $K_1$-point. The
 dimension of the sides of the square is 0.02\,$\pi/a_0$, where $a_0$ is the lattice constant . We show a relaxed and a
 2\%-strained configuration for $\vartheta=0$\dg \ and 20\dg.  Although our plots show
 only the vicinity of the $K_1$ point, the remaining plots can easily be constructed.
 Three of the six $K$-points are equivalent: This can be seen by apparent simple
 arguments: an arbitrary $K$-point, $K_i$ corresponds to a set of different of
 $K$-points in the adjacent Brillouin zones ($K_{i+2},K_{i+4}$) . The equivalence to
 the remaining three $K$-points follows by the $C_2$ rotation (or inversion), where
 $K_i$ maps to $K_{i+3}$.  Thus the contour plots of $K_i$ {$i$=1,3,5} are identical,
 the remaining contour plots $K_i$ {$i$=2,4,6} are found by inversion.

As can be seen, for the relaxed configuration the Dirac cones coincide with the $K$-point. For the
strained configuration at $\vartheta=0$\dg \ the crossing moves
along the $K_3-M_3$-line. Now the origin of the bands in Fig.~\ref{bild:GKMG}(a)
becomes clear: the line along $K_2-M_2$ cuts the Dirac cones away from the center,
resulting in a small opening of the bands. 

For the strained configuration at $\vartheta=20$\,\dg \ shown in Fig.~\ref{bild:ElContour}(c) the
Fermi crossing moves away from the zone-edge into the Brillouin zone.  
The question of whether or not a gap opens in graphene under small strains must be
answered by looking into the appropriate direction in reciprocal space. The tip of the
Dirac cones, according to our DFT calculations, lie on lines through $K$ that enclose
an angle of $2\vartheta$ and the $K-M$ line [see Fig.~\ref{bild:ElContour}(c)]. Looking at the band
structure along the strained high-symmetry directions, in contrast, corresponds to
cuts of the Dirac cones not centered at the Fermi level crossing, seemingly
suggesting an energy gap. 
This becomes important for strain in  directions other than
$\vartheta=0$\dg \ and $\vartheta=30$\dg, when the Fermi level does not coincide with
the Brillouin zone edges.

We now turn our attention to the changes in the vibrational spectrum when strain is
applied.  The vibrational bands show the general trend of softening under
strain. In contrast, the lowest-energy acoustic mode hardens. This mode, which
shows a quadratic dependence for $q \rightarrow 0$ for unstrained graphene, is an out-of-plane mode and
therefore referred to as ZA-mode\cite{mohr07gr}. The hardening only occurs for the
line in the $\Gamma-K_2$ direction. Also the quadratic dependence for $q \rightarrow
0$ changes into a linear dependence. This line describes propagating waves along the
direction of the applied strain. Thus the hardening can be compared to the frequency
increase when tension is applied to a string.

At the $\Gamma$-point the high-energy $E_{\mbox{\tiny2g}}$-phonon at 1581\,\wn \ is
two-fold degenerate for graphene. This degeneracy is lifted under strain.  As shown in
Ref.~\onlinecite{mohiuddin09} and Ref.~\onlinecite{huang09} the $E_{\mbox{\tiny
    2g}}$-mode of graphene splits into two distinct modes. These two modes possess
eigenvectors parallel and perpendicular to the strain direction. The mode parallel
(perpendicular) to the strain direction undergoes a larger (smaller) redshift and is
therefore entitled $G^-$ ($G^+$).  As discussed in 
Ref.~\onlinecite{mohiuddin09}  the strain rate of the $G^-$ ($G^+$) -mode is
independent on the direction of strain, a result which we find confirmed in our
calculations. This result stems from the isotropy of the hexagonal lattice.

\begin{figure}[htbp]
  \epsfig{file=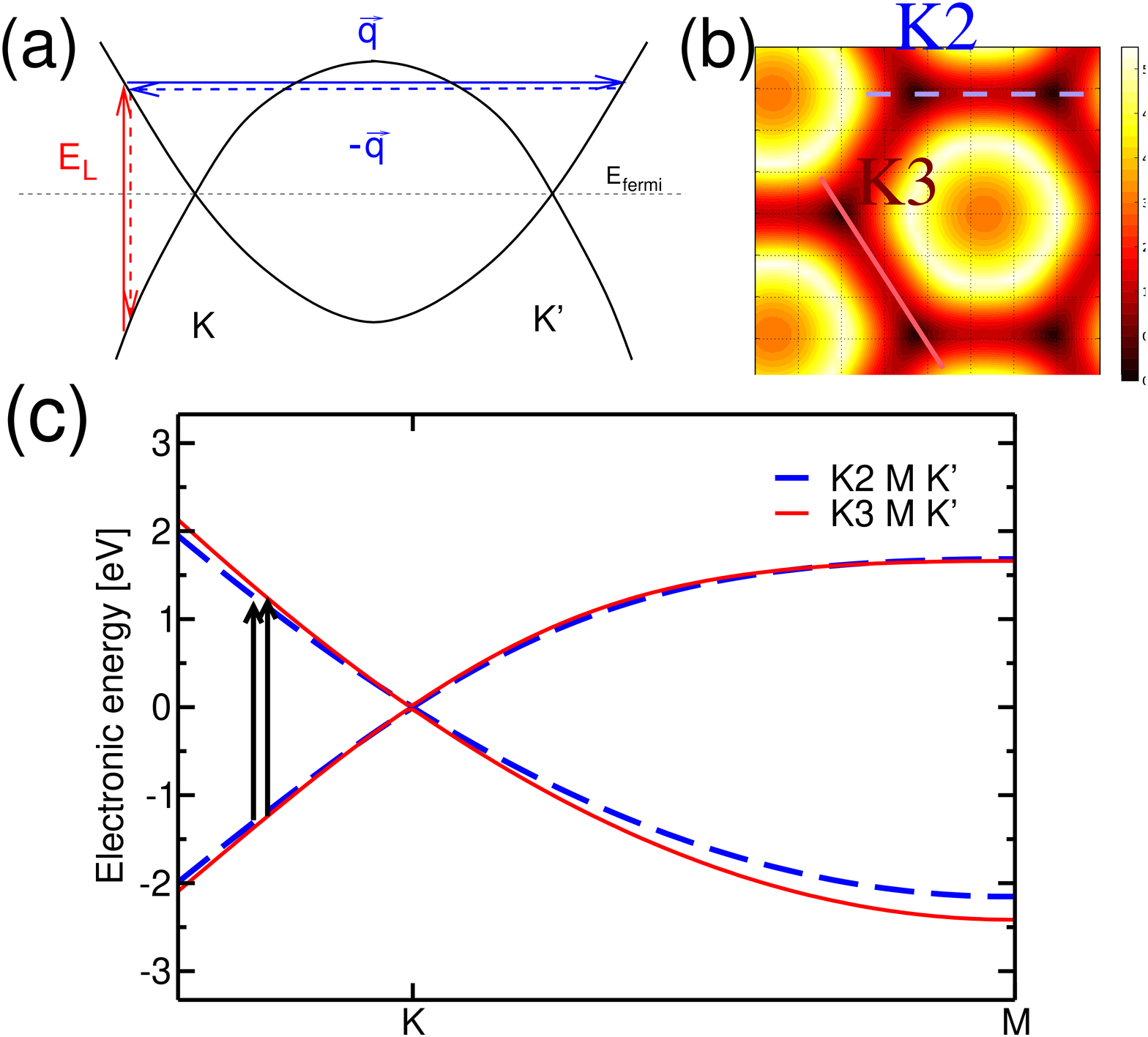,width=0.9\columnwidth}
 \caption{\label{bild:DR} (Color online) (a) Double resonance mechanism for the
   $2D$ mode. The energy of the scattering phonons is neglected. (b) Contour plot of the
   lowest-energy conduction band. (c) Band structure between $K$ and $K'$ between
   different $K$-points. The corresponding paths are shown in (b). Depending on the electronic
   transition a different wave vector of the scattered phonon becomes resonantly enhanced.
   This leads to a broadening of $2D$ mode, as contributions from the vicinity of all
   $K$-points are added up.
 }
 \end{figure}

In graphene the shape of the double-resonant $2D$ mode gives information on the number
of layers\cite{ferrari06}. This mode is energy-dependent and double
resonant\cite{thomsen00,maultzsch04dr}.  The mechanism of a DR-process is shown in
Fig.~\ref{bild:DR}(a). Here the phonon energy is assumed to be zero.  The contributing
phonon branch, the TO-mode,  is the mode with the highest energy between $K$ and
$M$ [see Fig. \ref{bild:GKMG}(b)].  Although the $2D$ mode contains contributions from
all over the Brillouin zone, the main contributions come from the $K-K'$ valleys
as shown by Narula and Reich\cite{narula08}.  Thus the one-dimensional treatment of
the DR gives a good approximation.

\begin{figure}[htbp]
  \epsfig{file=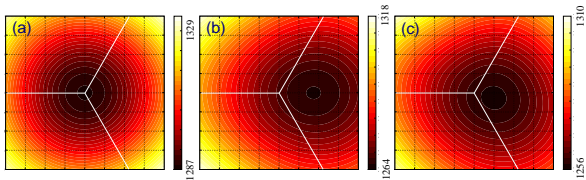, width=0.9\columnwidth}
  \caption{\label{bild:PhContour} (Color online) Contour plot of the fully symmetric
    TO phonon branch centered at the $K_1$-point. The TO mode contributes exclusively
    to the D-mode.  The dimension of the sides of the square is 0.2\,$\pi/a_0$ (a)
    Relaxed configuration, (b) 2\%-strained configuration for $\vartheta=0$\,\dg \ and
    (c) 2\%-strained configuration for $\vartheta=20$\,\dg.  }
 \end{figure}

The changes of the electronic bands and the vibrational bands both influence the
$2D$-mode.  Contour plots of the TO-mode energy are shown in Fig.~\ref{bild:PhContour}
for the relaxed and the 2\%-strained configuration for $\vartheta=0$\dg \ and
20\dg. Each plot is centered at its respective $K_1$-point.  Although the general
shape of the vibrational mode differs more strongly than the electronic bands under
strain, the stronger contribution for the shift of the $2D$ mode comes from changes in
the electronic structure.

Figure~\ref{bild:DR}(b) shows a contour plot of the lowest-energy conduction band and
two different $K-K'$ paths for the $\vartheta=$0\dg direction. In
Fig.~\ref{bild:DR}(c) we plot the electronic bands along these two paths. Depending on
the wave vector of the electronic transition a different phonon wave vector is doubly
resonant enhanced. This leads to differences in phonon energy of up to 10\wn /\%
strain and explains the broadening of the $2D$ mode under
strain\cite{mohiuddin09,huang09}. Experimentally the broadening is found to be around
13\wn /\% strain\cite{mohiuddin09}.

In Table~\ref{tab:rates} we show a summary of the determined shift rates for the
$G^+$, $G^-$ and $2D$ modes. These rates are for small strain values up to 2\,\%.  Our
calculated values are in excellent agreement with the experimental results from
Ref~.\cite{mohiuddin09}, although the calculated values are a bit higher.  In contrast to the $G^+$ and $G^-$-modes, the
$2D$ mode behaves slightly non-linear for strain greater than 1\%. Therefore we
suggest to measure the $G^+$ and $G^-$-modes for determining the strain rather
than the $2D$ mode.

In summary, we have presented an in-depth analysis of the electronic and vibrational
properties of uniaxially strained graphene. We demonstrated that graphene remains
semimetallic under small strain. The change of the Fermi surface suggests favored
directions for electronic transport depending on the direction of strain. Our calculated shift
rates of the Raman-active G and $2D$-band will help experimentalists to determine the
strain. Due to a deformation of the three-fold Dirac cone around the $K$-point the
double resonance condition changes and gives rise to a broadening of the $2D$ mode.

   \begin{table}
     \caption{\label{tab:rates}Shift rates for the $G^+$, the $G^-$ and 
the $2D$ mode in strained graphene (in \wn/\% strain). For the excitation energy dependent
$2D$ mode the excitation energy is written  parenthesis. }
     \begin{tabular}{c|c|c|c|c}
     Raman mode  & Ref.\cite{ni08}      &  Ref.\cite{mohiuddin09} & Ref.\cite{huang09} & this
       work \\ \hline
   $G^+$  &   -14.2 & -10.8  & -5.6  & -14.5 \\
   $G^-$  & -- \footnote{no distinction between $G^+$ and $G^-$  } & -31.7  & -12.5   & -34.0 \\
   $2D$ (2.41\,eV)      & -- &  -64  &--  &  -46..54 \\
   $2D$ (2.33\,eV)      & -27.8 & --  & -21 & -46..54   \\
   $2D$ (1.96\,eV)     & -- & --  & --  & -46..55 \\
     \end{tabular}
   \end{table}

JM and CT acknowledge support  by the Cluster of
Excellence 'Unifying Concepts in Catalysis' coordinated by the TU Berlin and
funded by DFG.

\end{document}